\documentclass[traditabstract]{aa} 
\usepackage{graphicx}
\usepackage{txfonts}
\usepackage{natbib}
\usepackage{tikz}

\usepackage{hyperref}
\hypersetup{colorlinks=true,citecolor=blue,linkcolor=blue,urlcolor=blue}

\graphicspath{{figures/}}

\newcommand{\e}{{\rm e}}
\newcommand{\ii}{{\rm i}}
\newcommand{\Pa}{{\rm P}_\alpha\!}
\newcommand{\Qa}{{\rm Q}_\alpha\!}
\newcommand{\Pb}{{\rm P}}
\newcommand{\Qb}{{\rm Q}}

\def\abs#1{\left\vert#1\right\vert}

\newcommand{\dop}[2]{{#1}\!\cdot\!{#2}}

\newcommand{\moy}[2]{#2}

\def\crm{\cr\noalign{\medskip}}

\def\m@th{\mathsurround=0pt}
\def\EQM#1{\vcenter{\normalbaselines\m@th
    \ialign{${\displaystyle ##}$\hfil&&\ ${\displaystyle ##}$\hfil\crcr
    \mathstrut\crcr\noalign{\kern-\baselineskip}
    \noalign{\smallskip}
    #1\crcr\mathstrut\crcr\noalign{\kern-\baselineskip}}}}

\newcommand{\be}{\begin{equation}}
\newcommand{\ee}{\end{equation}}

\def\Dron#1#2{\frac{\partial#1}{\partial#2}}

\newcommand{\bpm}{\left(\begin{array}{c}}
\newcommand{\epm}{\end{array}\right)}
\def\mybf#1{#1}

\newcommand{\figComp}{
\begin{figure}
\begin{tikzpicture}
\path(-4.5,0) node[rotate=90] {eccentricity};
\path(-2.5,1.0) node {a};
\path(0,0) node 
 {\includegraphics[width=0.9\linewidth,viewport=70pt 235pt 530pt 440pt,clip]{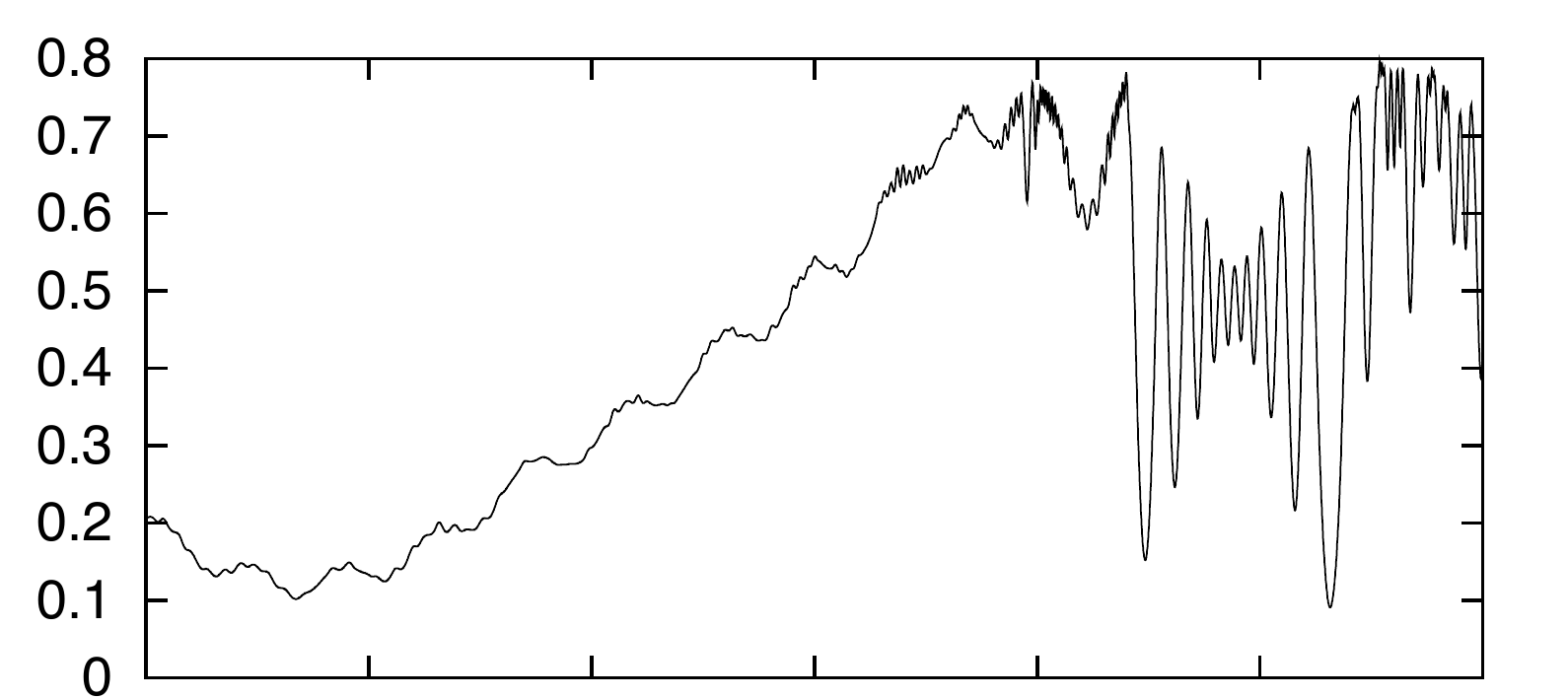}};
\path(-4.5,-3.8) node[rotate=90] {eccentricity};
\path(0,-4.1) node 
 {\includegraphics[width=0.9\linewidth,viewport=21pt 0pt 253pt 127pt,clip]{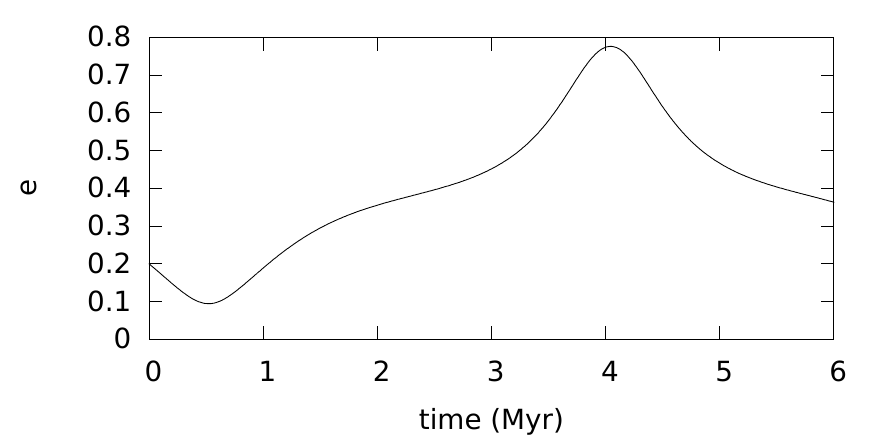}};
\path(-2.5,-2.7) node {b};
\end{tikzpicture}
\caption{\label{fig.Comp}Comparison between the eccentricity evolution
observed in \citep{Laskar_Icarus_2008} (a), and produced by the simple
model of this work (b).}
\end{figure}
}

\newcommand{\figfixpt}{
\begin{figure*}
\begin{center}
\includegraphics[width=0.75\linewidth]{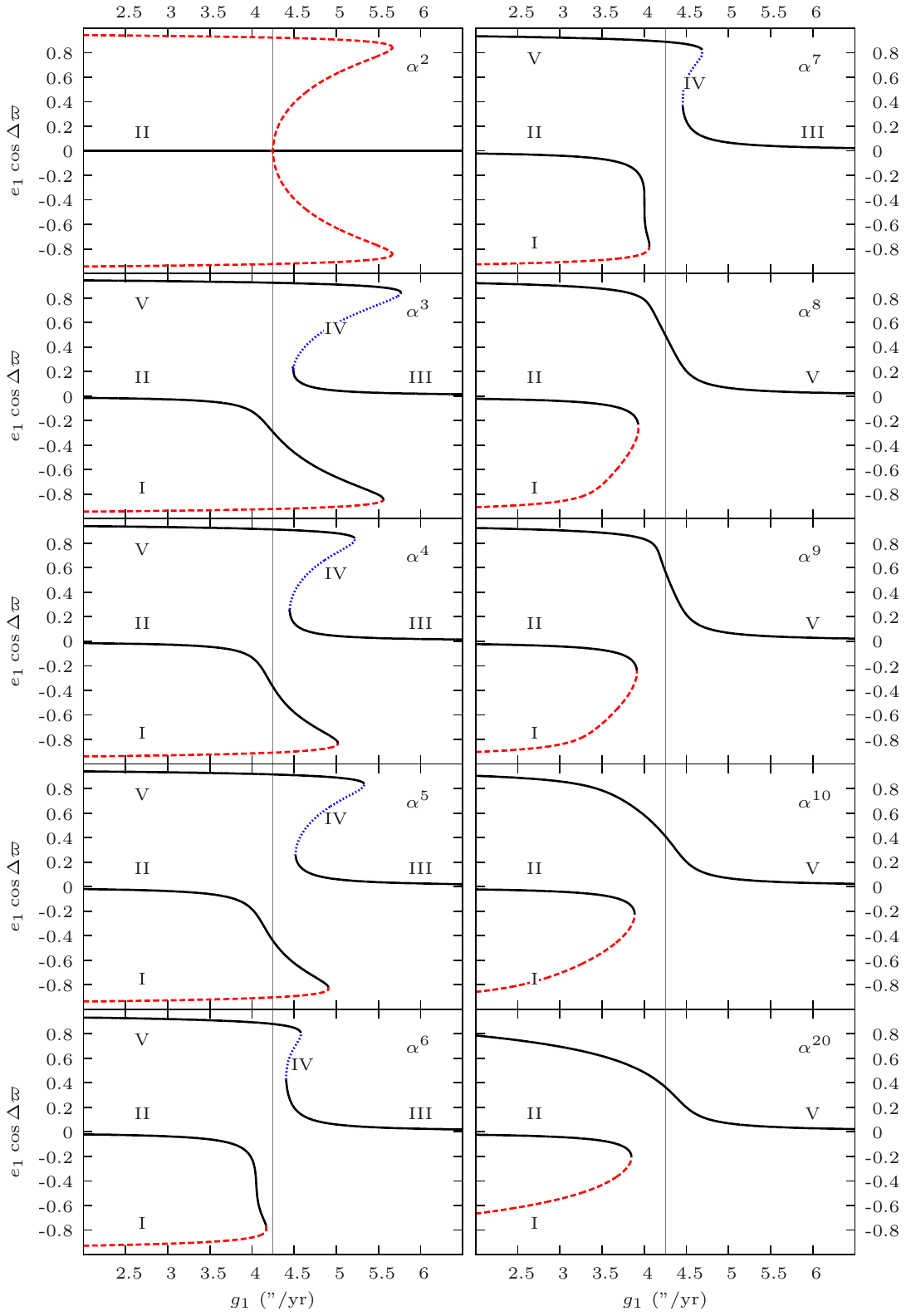}
\caption{\label{fig.fixpt} Evolution of the fixed points as a function
of the precession frequency of Mercury's orbit at zero eccentricity and
for different expansion orders. Dashed and dotted curves correspond to
unstable points, while the solid ones show the positions of the stable
points. \mybf{The equilibrium points are labeled with roman letters as in
Fig.~\ref{fig.section}.} The thin vertical line indicates the resonance $g_1=g_5$.}
\end{center}
\end{figure*}
}

\newcommand{\figsection}{
\begin{figure*}
\begin{center}
\includegraphics[width=0.8\linewidth]{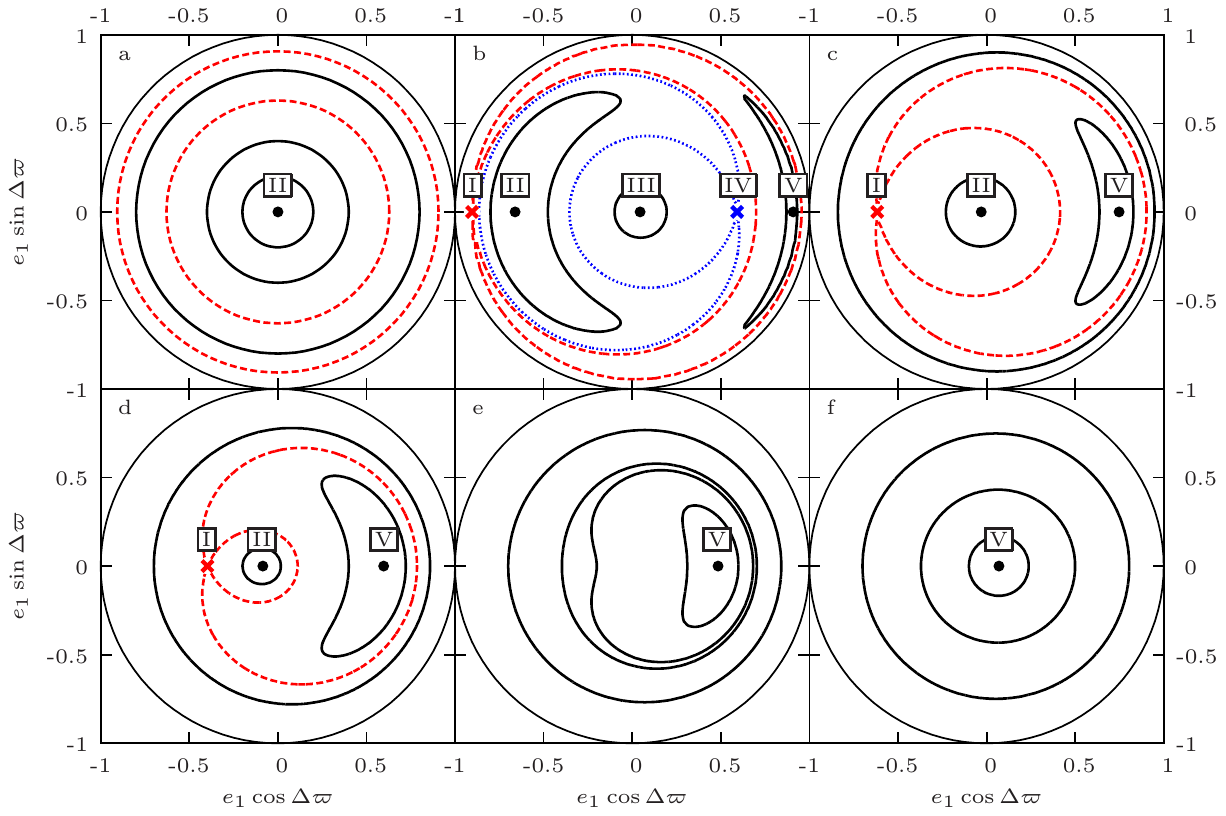}
\caption{\label{fig.section} Level curves of the Hamiltonian in the
plane $(e_1\cos\Delta\varpi, e_1\sin\Delta\varpi)$. a) Quadrupole
expansion with $g_1=5$"/yr. \mybf{b) Octupole expansion with
$g_1=5$"/yr.  c,d,e,f) expansion up to the order $n=20$ with
$g_1=2.5,3.6,4.0,5.0$"/yr, respectively.}
}
\end{center}
\end{figure*}
}

\newcommand{\fignsection}{
\begin{figure}
\begin{center}
\includegraphics[width=0.5\linewidth]{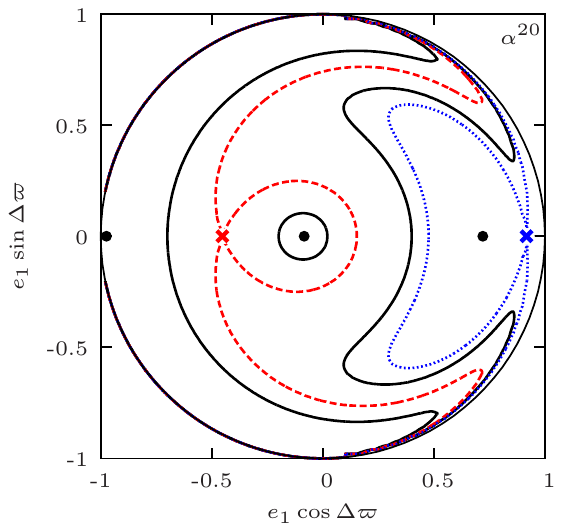}%
\includegraphics[width=0.5\linewidth]{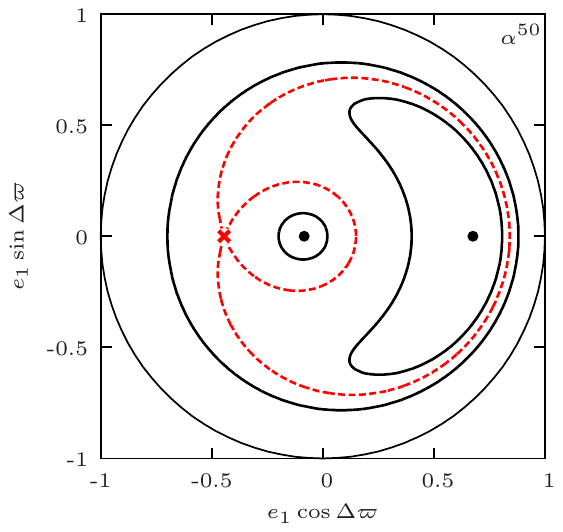}
\caption{\label{fig.nsection} Level curves of the Hamiltonian without
relativity in the plane $(e_1\cos\Delta\varpi, e_1\sin\Delta\varpi)$ for
two orders of expansion: $n=20$ (left), $n=50$ (right). In both cases,
$g_1=3.6$"/yr.}
\end{center}
\end{figure}
}

\newcommand{\fignewt}{
\begin{figure}
\begin{center}
\includegraphics[width=1.0\linewidth]{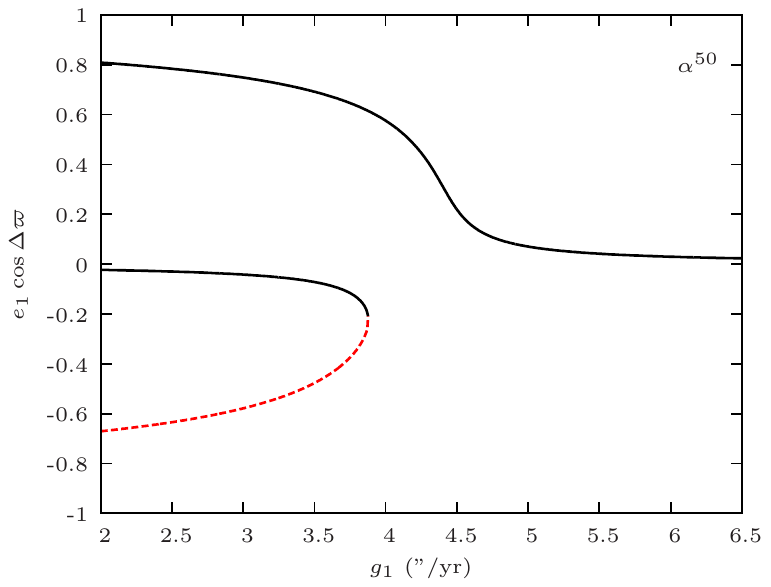}
\includegraphics[width=1.0\linewidth]{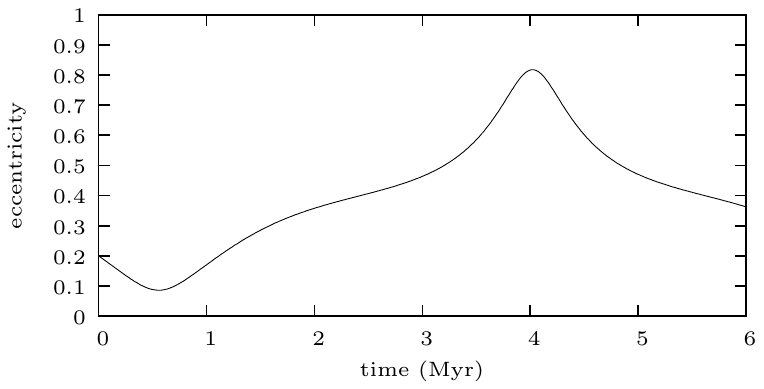}
\caption{\label{fig.newton} Top: evolution of the fixed points of $H^{(50)}$ 
without relativity. Bottom: evolution of the eccentricity with time,
using the simple model without relativity.}
\end{center}
\end{figure}
}

\newcommand{\figincli}{
\begin{figure}
\begin{center}
\includegraphics[width=\linewidth]{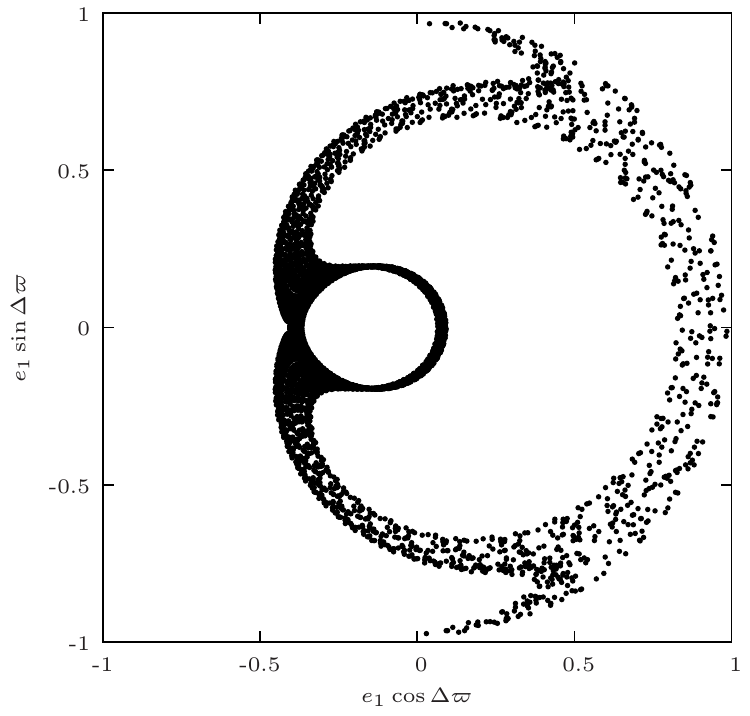}
\includegraphics[width=\linewidth]{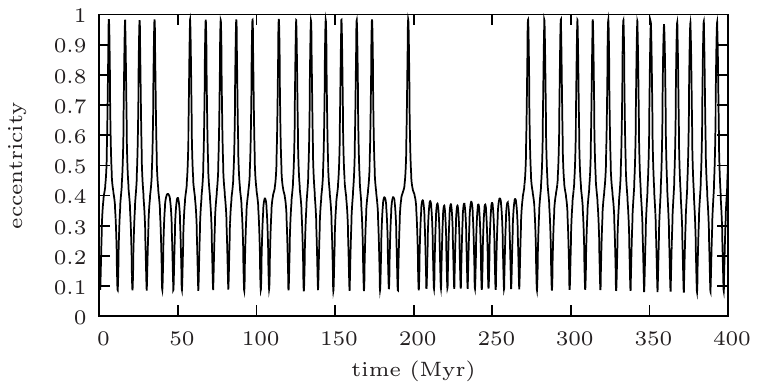}
\caption{\label{fig.incli} Top: trajectory of Mercury's eccentricity in
the plane $(e_1\cos\Delta\varpi, e_1\sin\Delta\varpi)$ for the inclined
system. Bottom: evolution of Mercury's eccentricity as a function of
time in the same system.}
\end{center}
\end{figure}
}

\newcommand{\tabAmp}{
\begin{table}
\begin{center}
\renewcommand{\arraystretch}{1.5}
\caption{\label{tab.amp} Amplitude of the quadrupole $\varepsilon_p^{\rm quad}$ 
(\ref{eq.eps_quad}) and octupole $\varepsilon_p^{\rm octu}$ (\ref{eq.eps_octu}) 
terms, 
due to each planet $p$, amplitude $A_p$, and phase $\varphi_p$ of the eigenmode with 
frequency $g_5=4.2488163$"/yr in the quasiperiodic decomposition of 
$z_p = e_p \exp i\varpi_p$ \citep[taken from][]{Laskar_Icarus_1990}.}
\begin{tabular}{*{5}r} \hline
$p$ & $\varepsilon_p^{\rm quad} \times 10^6$
    & $\varepsilon_p^{\rm octu} \times 10^6$ & $A_p\times 10^6$ & $\varphi_p$ (deg.) \\
\hline\hline
2   &  59\,375 & 1\,170 & 19\,636 &  30.571 \\
3   &  27\,911 &    383 & 18\,913 &  30.597 \\
4   &      837 &      8 & 20\,300 &  30.679 \\
5   &  62\,201 &    383 & 44\,119 &  30.676 \\
6   &   3\,025 &      8 & 33\,142 &  30.676 \\
7   &       57 &      0 & 37\,351 & 210.671 \\
8   &       17 &      0 &  1\,771 &  30.669 \\
\hline
\end{tabular}
\end{center}
\end{table}
}

\newcommand{\tabBa}{
\begin{table}
\begin{center}
\renewcommand{\arraystretch}{1.2}
\caption{\label{tab.B1} Coefficients of the polynomials $\Pb(x) = \sum p_\ell x^\ell$ 
(\ref{eq.Pbnum}) and $\Qb(x) = \sum q_\ell x^\ell$ (\ref{eq.Qbnum})
computed up to $\alpha^{50}$.}
\begin{tabular}{*{3}r|*{3}r} \hline
$\ell$ & $p_\ell \times 10^8$ & $q_\ell \times 10^8$ &
$\ell$ & $p_\ell \times 10^8$ & $q_\ell \times 10^8$\strut \\
\hline\hline
  0 &$ 33658\,599 $&$  1215\,100 $& 13 &$ 7651\,739 $&$ 4402\,899 $  \\
  1 &$ 65113\,787 $&$  2456\,366 $& 14 &$ 4380\,363 $&$ 2117\,788 $  \\
  2 &$ 19186\,677 $&$  2568\,036 $& 15 &$ 1965\,774 $&$  775\,772 $  \\
  3 &$ 12894\,952 $&$  2963\,700 $& 16 &$  674\,911 $&$  211\,606 $  \\
  4 &$ 10843\,781 $&$  3544\,094 $& 17 &$  173\,233 $&$   41\,954 $  \\
  5 &$ 10208\,736 $&$  4322\,435 $& 18 &$   32\,433 $&$    5\,875 $  \\
  6 &$ 10268\,763 $&$  5332\,153 $& 19 &$    4\,303 $&$       560 $  \\
  7 &$ 10780\,251 $&$  6598\,519 $& 20 &$       390 $&$        35 $  \\
  8 &$ 11612\,770 $&$  8057\,388 $& 21 &$        23 $&$         1 $  \\
  9 &$ 12557\,438 $&$  9404\,855 $& 22 &$  0.8      $&$ 0.03      $  \\
 10 &$ 13156\,622 $&$ 10023\,739 $& 23 &$  0.02     $&$ 0.000\,2  $  \\
 11 &$ 12724\,433 $&$  9279\,738 $& 24 &$  0.000\,1 $&$0.000\,0006$  \\
 12 &$ 10785\,258 $&$  7137\,951 $& 25 &$0.000\,0003$&         --    \\
\hline
\end{tabular}
\end{center}
\end{table}
}

\newcommand{\tabBb}{
\begin{table}
\begin{center}
\renewcommand{\arraystretch}{1.2}
\caption{\label{tabB2} Frequencies at zero eccentricity and corrections 
for different orders of expansion of the perturbing function.}
\begin{tabular}{r*2{|rr}} 
\multicolumn{1}{c}{}
& \multicolumn{2}{ c }{Newtonian case} & 
  \multicolumn{2}{c }{Relativistic case} \\ \hline 
$n$ \rule[-0.3\baselineskip]{0pt}{1.2em} & 
      $g_1$ ("/yr) & 
      $\delta_g^{(r)}$ ("/yr) & 
      $g_1$ ("/yr) & 
      $\delta_g^{(r)}$ ("/yr) \\ \hline\hline
 2 & 3.91118 &  0.33764 & 4.32283 & -0.07401 \\ 
 3 & 3.91118 &  0.33764 & 4.32283 & -0.07401 \\ 
 4 & 4.94380 & -0.69499 & 5.35546 & -1.10664 \\ 
 5 & 4.94380 & -0.69499 & 5.35546 & -1.10664 \\ 
 6 & 5.32808 & -1.07926 & 5.73973 & -1.49092 \\ 
 7 & 5.32808 & -1.07926 & 5.73973 & -1.49092 \\ 
 8 & 5.46485 & -1.21603 & 5.87650 & -1.62769 \\ 
 9 & 5.46485 & -1.21603 & 5.87650 & -1.62769 \\ 
10 & 5.51200 & -1.26318 & 5.92365 & -1.67483 \\ 
11 & 5.51200 & -1.26318 & 5.92365 & -1.67483 \\ 
12 & 5.52788 & -1.27907 & 5.93954 & -1.69072 \\ 
13 & 5.52788 & -1.27907 & 5.93954 & -1.69072 \\ 
14 & 5.53314 & -1.28433 & 5.94480 & -1.69598 \\ 
15 & 5.53314 & -1.28433 & 5.94480 & -1.69598 \\ 
16 & 5.53486 & -1.28604 & 5.94651 & -1.69770 \\ 
17 & 5.53486 & -1.28604 & 5.94651 & -1.69770 \\ 
18 & 5.53541 & -1.28659 & 5.94706 & -1.69825 \\ 
19 & 5.53541 & -1.28659 & 5.94706 & -1.69825 \\ 
20 & 5.53559 & -1.28677 & 5.94724 & -1.69842 \\ 
21 & 5.53559 & -1.28677 & 5.94724 & -1.69842 \\ 
22 & 5.53564 & -1.28683 & 5.94730 & -1.69848 \\ 
23 & 5.53564 & -1.28683 & 5.94730 & -1.69848 \\ 
24 & 5.53566 & -1.28684 & 5.94731 & -1.69850 \\ 
25 & 5.53566 & -1.28684 & 5.94731 & -1.69850 \\ 
26 & 5.53567 & -1.28685 & 5.94732 & -1.69850 \\ 
   & \multicolumn{2}{c|}{- - -} & \multicolumn{2}{c}{- - -} \\
50 & 5.53567 & -1.28685 & 5.94732 & -1.69851 \\ \hline

\end{tabular}
\end{center}
{\bf Notes:} $g_1$ is Mercury's precession frequency computed at 
$e_1=0$ without the correction $\delta_g e_1^2/2$ in each Hamiltonians.
$\delta_g^{(r)}$ represents the correction $\delta_g$ that puts the
system in resonance ($g_1=g_5$).
\end{table}
}

\begin{document}
   \title{A simple model of the chaotic eccentricity of Mercury}

   \subtitle{}

   \author{Gwena\"el Bou\'e\inst{1,2,3}
           \and
           Jacques Laskar\inst{2}
           \and
           Fran\c{c}ois Farago\inst{2}
          }

   \institute{
             Centro de Astrof\'isica, Universidade do Porto, Rua das
             Estrelas, 4150-762 Porto, Portugal
         \and
             Astronomie et Syst\`emes Dynamiques, IMCCE-CNRS UMR8028,
             Observatoire de Paris, UPMC, 77 Av. Denfert-Rochereau, 
             75014 Paris, France
         \and
             Department of Astronomy and Astrophysics, University of
             Chicago, 5640 South Ellis Avenue, Chicago, IL 60637, USA\\
             \email{boue@oddjob.uchicago.edu}
             }

   \date{Received \ldots,\ldots; accepted \ldots,\ldots}

 

  \abstract
{Mercury's eccentricity is chaotic and can  increase so much that collisions
with Venus or the Sun become  possible.
 This chaotic behavior 
results from an intricate network of secular resonances, but in this paper, we show that a simple
integrable model with only one degree of freedom is actually able to 
reproduce the large variations in Mercury's eccentricity, with the correct 
amplitude and timescale. We show that this behavior occurs in the vicinity of the separatrices  of the
resonance $g_1-g_5$ between the precession frequencies of Mercury 
and Jupiter. However, the main contribution does not come from the
direct interaction between these two planets. It is due to the
excitation of Venus' orbit at Jupiter's precession frequency $g_5$.
\mybf{We use a multipolar model that is not expanded with respect to Mercury's eccentricity, but because of the proximity of Mercury and Venus, 
the Hamiltonian is expanded up to order 20 and more in the ratio of semimajor axis.}
When  the effects of Venus' inclination are added, the system becomes
 nonintegrable and a chaotic zone appears in the vicinity of the separatrices. 
 In that case, Mercury's eccentricity can chaotically
switch between two regimes characterized by either low-amplitude
circulations or high-amplitude librations.
}


   \maketitle
%

\section{Introduction}
After the discovery of the chaotic motion of the planets in the  Solar System 
\citep{Laskar_Nature_1989, Laskar_Icarus_1990}, it has been demonstrated that 
the eccentricity of Mercury can rise to very high values  
\citep{Laskar_AA_1994}, allowing for collision of the planet with Venus. 
This was confirmed later on by direct numerical integration 
in simplified models that neglect general relativity (GR)
\citep{Laskar_Icarus_2008, Batygin_Laughlin_ApJ_2008}, 
and even in the full model that includes the GR contribution   
\citep{Laskar_Gastineau_Nature_2009}.
Quite surprisingly, the behavior of the system depends strongly on the 
GR contribution. Indeed, the probability that the eccentricity 
increases beyond 0.7 in less than 5 Gyr is about 1\% in the full model, 
while it rises to more than 60\% when GR is neglected 
 \citep{Laskar_Icarus_2008, Laskar_Gastineau_Nature_2009}.
 This behavior is due to the presence of a secular resonance 
 between the perihelion motions of Mercury ($\varpi_1$) and Jupiter ($\varpi_5$)
  \citep{Laskar_Icarus_2008, Batygin_Laughlin_ApJ_2008}.
Although the origin of the instability in Mercury's eccentricity 
is identified, the precise mechanism of the behavior of 
Mercury's orbit has not yet been explained in detail. A first attempt 
was provided by \citep{Lithwick_Wu_ApJ_2011}, who analyzed the 
overlap of   resonances in the truncated secular Hamiltonian of
degree four in eccentricity and inclination. This model reproduces
the small diffusion of Mercury's eccentricity well,
and confirms the important role of the $g_1-g_5$ and $(g_1-g_5)-(s_1-s_2)$ 
secular resonances between the  precession frequency of the perihelion ($g_i$) 
and the regression rate of the ascending node $(s_i)$.  
However, \citet{Lithwick_Wu_ApJ_2011} does
not explain the steady increase in Mercury's eccentricity beyond 0.7 as
observed in \citep{Laskar_Icarus_2008}.

In the present  paper, our approach is different, 
since we do not want to be bound to the limitations given by expansions 
in eccentricity as in \citep{Lithwick_Wu_ApJ_2011} or in the 
previous work of \citet{laskar_theorie_1984,Laskar_Nature_1989, Laskar_Icarus_1990, Laskar_Icarus_2008}.
As we are looking for a large excursion of the eccentricity of Mercury, 
we prefer to derive expressions that are valid for all 
eccentricities, using the averaged expansions derived in
\citep{Laskar_Boue_AA_2010}. We provide \mybf{two simple} analytical,
\mybf{secular} models\mybf{. The first one is coplanar with only one degree of freedom.
It is thus  integrable. This model shows that Mercury's eccentricity
can reach values as high as 0.8 if the system is in the vicinity of
the $g_1-g_5$ resonance. The second model includes inclination and has
two degrees of freedom. This spatial model exhibits the chaotic behavior of Mercury's
eccentricity in the neighborhood of the separatrix with possibilities
of switching between a regime of moderate eccentricity to a regime of  large
oscillation where Mercury's  eccentricity reaches very high values. In both models, 
} 
Mercury is treated as a massless particle, while the motion of the other
planets in the Solar System are given by a single term in their
quasiperiodic decompositions taken from \citep{Laskar_Icarus_1990}. In
simplified terms, the chaos in Mercury's orbit is due to the proximity
of the resonance with the precession motion of Venus excited at the
frequency of Jupiter's precession frequency.

The secular models that we use in this paper rely on a multipolar
expansion of the perturbing function, up to order $(a_1/a_p)^{20}$ in
the relativistic case and $(a_1/a_p)^{50}$ in the Newtonian case, 
where $(a_p)_{p=2,8}$ are the semimajor axes of the planets of the Solar
System in increasing order. This expansion, subsequently averaged over
the mean anomalies of all the planets, allows for arbitrary
inclinations and eccentricities as long as no orbit crossing occurs.

In section 2, we derive the equations of motion \mybf{of the coplanar model}. This model is very
simple with only one degree of freedom associated to Mercury's
eccentricity and longitude of perihelion. Then, in section 3, we derive
the possible trajectories in the phase space using level curves of the
Hamiltonian for different orders of expansion. We show that it is
necessary to develop the perturbing function up to high orders in
$(a_1/a_p)$ in order to reach the asymptotic evolution. We also show
that the maximum eccentricity attained with an initial eccentricity of
$e_1=0.2$ is on the order of 0.8 within 4 Myrs as reported in
\citep{Laskar_Icarus_2008}. The \mybf{ spatial case is treated in section 4. 
This nonintegrable model illustrates the generation of chaos in the 
vicinitiy of the hyperbolic fixed point of the  planar model.} We conclude in the last section.

\section{Coplanar model}
\label{planar}
\figsection

\figfixpt
\subsection{Newtonian interaction}
Using the traditional notations where planets have increasing indices
with respect to their semimajor axis, we note the barycentric position
of the Sun $\vec u_0$, of Mercury $\vec u_1$, of Venus $\vec u_2$, etc.
The heliocentric positions of the planets are similarly noted $(\vec
r_p)_{p=1,8}$. Since the mass of Mercury is much lower than the masses
of the other planets in the Solar System, we model Mercury as a massless
particle. 

Our goal is to study the behavior of Mercury under a
known planetary perturbation. As such, all of the $\vec u_p$ and their
derivatives, for $p\neq 1$, are considered as given functions of the
time $t$. Using the Poincar\'e heliocentric canonical variables \citep{Laskar_Robutel_CeMe_1995}, the Hamiltonian
describing the evolution of Mercury's trajectory reads as\footnote{For 
clarity, we drop the explicit time dependence in all equations
after (\ref{eq.Htime}).}
\be
\hat{H}_N = \frac{\tilde{\vec r}_1^2}{2} 
  - G \frac{m_0}{r_1} 
  - G\sum_{p=2}^8 \frac{m_p}{\abs{\vec r_1 - \vec r_p(t)}}
  +  \sum_{p=2}^8 \frac{m_p}{m_0}{\dop{\tilde{\vec r}_1}{\dot{\vec u}_p(t)}}\ ,
\label{eq.Htime}
\ee
where $\tilde{\vec r}_1 = \dot{\vec u}_1$ is the conjugate momentum of
$\vec r_1$, $G$ the gravitational constant, $m_0$ the
mass of the Sun, and $(m_p)_{p=2,8}$ the masses of the other planets.
The first two terms of this Hamiltonian represent the Keplerian motion
and are equal to $-Gm_0/(2a_1)$, where $a_1$ is the semimajor axis of
Mercury.

Considering that Mercury is the closest planet to the Sun, we
now expand $\hat{H}_N$ formally as a series in $(r_1/r_p)_{p=2,8}$
\be
\hat{H}_N = -\frac{Gm_0}{2a_1} 
   -G\sum_{p=2}^8 \frac{m_p}{r_p} \sum_{n=0}^{\infty} 
    \left(\frac{r_1}{r_p}\right)^n P_n 
           \left(\frac{\dop{\vec r_1}{\vec r_p}}{r_1 r_p}\right)
  +  \sum_{p=2}^8 \frac{m_p}{m_0}{\dop{\tilde{\vec r}_1}{\dot{\vec u}_p}}\ ,
\label{eq.HamPl}
\ee
where $P_n$ is the Legendre polynomial of order $n$.

A first approximating model could stop the expansion at the second
Legendre polynomial $P_2$. However, it is well known that the resulting
double-averaged secular Hamiltonian does not depend on the perihelia of
the outer bodies at this order \citep{Lido_1962_a,Kozai_AJ_1962,
Lidov_Ziglin_CeMe_1976, Farago_Laskar_MNRAS_2010}, so no secular
resonance between the perihelia of Mercury and any other planet can be 
seen there. We thus push the expansion to higher orders.
Furthermore, we see in the following that it is necessary to extend
the sum at least up to $n \approx 10$. 

The next step consists in averaging the Hamiltonian (\ref{eq.HamPl})
over the mean anomalies $(M_p)_{p=1,8}$ of all planets. In the
Hamiltonian (\ref{eq.HamPl}), 
the terms $-Gm_0/r_p$ obtained for $n=0$ do not depend on Mercury's elements;
the terms $({\tilde{\vec r}_1} \cdot {{\vec u}_p})$
vanish when averaged over $M_1$; the term $-Gm_0/(2a_1)$ becomes 
constant after averaging over $M_1$, since $a_1$ becomes constant. As such,
all these terms are dropped in the following expressions. Finally, the
averaged expression of the perturbing functions $a_p/\abs{\vec r_1-\vec
r_p}$ are given in \citep{Laskar_Boue_AA_2010}. The resulting
Hamiltonian of the coplanar problem is then
\be
\moy{M_p}{\hat {H}_{N,{\rm plan}}} = -G\sum_{p=2}^8 \frac{m_p}{a_p} \sum_{n=2}^\infty 
\left(\frac{a_1}{a_p}\right)^n
{\cal F}_n^{(0,0)} (e_1, e_p, \varpi_1-\varpi_p)\ ,
\label{eq.Hsec}
\ee
where
\be
\EQM{
{\cal F}_n^{(0,0)} (e_1, e_p, \varpi) = 
\epsilon_n f_{n,\frac{n}{2}} X_0^{n,0}(e_1) X_0^{-(n+1),0} (e_p) \cr
+ \sum_{q=0}^{[(n-1)/2]} 2f_{n,q} X_0^{n,n-2q}(e_1) X_0^{-(n+1),n-2q}(e_p)
\cos((n-2q)\varpi)\ .
\label{eq.Fn00}
}
\ee
In these expressions, $a_p$, $e_p$, and $\varpi_p$ are the
semimajor axis, the eccentricity, and the longitude of the pericenter of
the planet $p$, respectively. The $X_k^{n,m}(e)$ are the Hansen coefficients defined by
\be
\left(\frac{r}{a}\right)^n \e^{imv} = \sum_{k=-\infty}^{\infty}
X_{k}^{n,m}(e) \e^{i k M}\ ,
\ee
where $\epsilon_n=0$ if $n$ is odd, and $\epsilon_n=1$ if $n$
is even, and
\be
f_{n,q} = \frac{(2q)!(2n-2q)!}{2^{2n}(q!)^2((n-q)!)^2}\ .
\ee

To simplify the expansion of the Hamiltonian (\ref{eq.HamPl}), we take
advantage of the eccentricities $(e_p)_{p=2,8}$ of all the planets
beyond Mercury remaining low to only keep the linear terms in these
eccentricities. In contrast, the expansion remains exact in the
eccentricity $e_1$. With this approximation, up to the {\em octupole}
order, the averaged Hamiltonian (\ref{eq.HamPl}) reads as
\be
\EQM{
\moy{M_p}{\hat {H}_{N, {\rm plan}}}
\approx -\frac{G}{8}\sum_{p=2}^8  & m_p 
\bigg(\frac{a_1^2}{a_p^3} (2+3e_1^2)
\cr &
-\frac{15}{8}\frac{a_1^3}{a_p^4} e_1 e_p (4+3e_1^2)                   
\cos(\varpi_1-\varpi_p) \bigg)\ .
}
\ee
More generally, as long as the Hamiltonian is truncated at the first
order in $e_p$, its expression can be written as (see
Appendix~\ref{sec.appplan})
\be
\EQM{
\moy{M_p}{\hat {H}_{N, {\rm plan}}} = -G \sum_{p=2}^8 \frac{m_p}{a_p} 
\bigg( & \Pa\left(\frac{a_1}{a_p},e_1^2\right) 
\cr &
- e_1 e_p \Qa\left(\frac{a_1}{a_p}, e_1^2\right) \cos
(\varpi_1-\varpi_p)\bigg)\ ,
}
\label{eq.HamPQ}
\ee
where $\Pa(\alpha, x)$ and $\Qa(\alpha, x)$ are polynomials 
of degree $[n/2]$ and $[(n-1)/2]$ in $x$ and of degree $2[n/2]$
and $2[(n+1)/2]-1$ in $\alpha$, with $n$ the degree of
expansion of the perturbing functions as in Eq.~(\ref{eq.Hsec}). It
can be noted that the coefficient of $x^0$ in $\Pa(\alpha,x)$ is the
Taylor expansion of $C_1(\alpha) -1 = (1/2) b_{1/2}^{(0)}(\alpha)$ -1,
and the coefficient of $x^1$ in $\Pa(\alpha, x)$ is the Taylor series of
$C_3(\alpha)/2 = (1/8)\alpha b_{3/2}^{(1)}(\alpha)$, etc., where
$b_{s}^{(k)}(\alpha)$ are Laplace coefficients
\citep{Laskar_Robutel_CeMe_1995}.

In the Hamiltonian (\ref{eq.HamPQ}), $e_p$ and $\varpi_p$ are functions 
of time. Neglecting the slow diffusion of the eccentricities of the
inner planets, the evolution of each $z_p = e_p \exp i\varpi_p$ is
described well by a quasiperiodic expansion \citep{Laskar_Icarus_1990}. In
this study, we focus on the secular resonance $g_1-g_5$. In
the quasiperiodic decomposition of the variables $(z_p)_{p=2,8}$, we
thus keep only the terms associated to the eigenmode $z_5^\star=\exp
i g_5 t$ with frequency $g_5\approx 4.25$"/yr. The others average
out and disappear from the Hamiltonian. 

\tabAmp

The amplitude and the phase of the mode $z_5^\star$ in the decomposition
of each variable $z_p$ are provided in Table~\ref{tab.amp}. We notice
that, except for Uranus ($p=7$), all the phases are the same and equal
to $\approx30.6\deg$. Since, $\varphi_7=30.6+180\deg$, it is equivalent to
take $\varphi'_7=30.6\deg$ and $A'_7=-A_7$. This is the convention that
is followed hereafter, but the prime is
omitted for clarity. We note $\varpi_5^\star=g_5 t + \varphi_5^\star$, where 
$\varphi_5^\star=30.6\deg$. \mybf{We also use the fact that 
$\Lambda_1=\sqrt{G m_0 a_1}$ is constant in the secular problem to
rescale the Hamiltonian by this quantity as it simplifies the 
following expressions. We note
$
\check{H}_{N,{\rm plan}} = 
\moy{M_p}{\hat{H}_{N,{\rm plan}}}/\Lambda_1\ .
$
}
Given that
$
Gm_p/(\Lambda_1 a_p) = n_1 a_1 m_p / (a_p m_0) \ ,
$
the resonant Hamiltonian now reads as
\be
\EQM{
\check{H}_{N,{\rm plan}} = -n_1 \sum_{p=2}^8 & \frac{m_p}{m_0}\frac{a_1}{a_p} 
\bigg( \Pa\left(\frac{a_1}{a_p},e_1^2\right) 
\cr &
- e_1 A_p \Qa\left(\frac{a_1}{a_p}, e_1^2\right) \cos
(\varpi_1-\varpi_5^\star)\bigg)\ .
}
\label{eq.Hamres0}
\ee
\mybf{
With this convention, the time should also be rescaled by the same
factor $\Lambda_1$ to keep the canonical form of the equations of motion, unless the
canonical variables are modified as follows. We let
\be
\EQM{
\hat I = \Lambda_1 \left(1-\sqrt{1-e_1^2}\right)\ , \crm
\hat \theta = -\varpi_1
}
\ee
be the canonical conjugated variables of the initial Hamiltonian $\moy{M_p}{\hat{H}_{N,{\rm
plan}}}$ of (Eq.\ref{eq.HamPQ}). It can be easily shown that
\be
(\check{I},\check{\theta}) = \left(\frac{\hat I}{\Lambda_1}, \hat \theta\right)
\ee
are canonical conjugated variables of the Hamiltonian $\check{H}_{N,{\rm plan}}$ (Eq.\ref{eq.Hamres0}) 
without rescaling the time $t$.
}
The equations of motion are thus
\be
\EQM{
\frac{d\check{I}}{dt} &= 
-\Dron{\check{H}_{N,{\rm plan}}}{\check \theta}
(\check I,\check \theta,t)\ ,
\qquad & 
\frac{d\check \theta}{dt} &= 
 \Dron{\check{H}_{N,{\rm plan}}}{\check I}
(\check I,\check \theta,t)\ .
}
\ee

At this point, it is interesting to evaluate the contribution of each
planet to the octupole interaction with Mercury. Although the resonance
$g_1-g_5$ involves only the precession frequencies associated to Mercury and
Jupiter, the eigenmode $z_5^\star$ is present in 
the quasiperiodic decomposition of the motion of all the planets.
Furthermore, the amplitudes of this mode are very similar and vary
only within a factor 2.5 between 0.019 and 0.044, except for Neptune
whose amplitude is only 0.002 (Table~\ref{tab.amp}). From the
expression of the Hamiltonian (\ref{eq.Hamres0}), knowing that the lowest
degree of $Q(\alpha, x)$ in $\alpha$ is 3, the contribution of the
octupole terms can be estimated using the parameters
$(\varepsilon^{\rm octu}_p)_{p=2,8}$ given by
\be
\varepsilon_p^{\rm octu} = 
\frac{15}{64}\left(\frac{n_1}{g_5}\right)\left(\frac{m_p}{m_0}\right)
\left(\frac{a_1}{a_p}\right)^4 A_p\ .
\label{eq.eps_octu}
\ee
The values taken by these parameters are gathered in Table~\ref{tab.amp}.
The maximal amplitude is due to Venus with $\varepsilon_2^{\rm
octu}\approx 0.0012$ followed by the Earth-Moon barycenter and Jupiter with 
$\varepsilon_3^{\rm octu}\approx \varepsilon_5^{\rm octu}\approx 4\times10^{-4}$.
Thus, the strongest perturbation on Mercury's orbit comes from the
precession of Venus excited by Jupiter.
$a_1/a_2\approx0.53$ is not very small  explains why it is
necessary to perform the expansion of the perturbing function up to a
high order in the ratio of the semimajor axes. For completeness,
Table~\ref{tab.amp} also provides the quadrupole contribution of each
planet given by
\be
\varepsilon_p^{\rm quad} = 
\frac{1}{8}
\left(\frac{n_1}{g_5}\right)\left(\frac{m_p}{m_0}\right)
\left(\frac{a_1}{a_p}\right)^3\ .
\label{eq.eps_quad}
\ee
Since we have shown that several planets play a role in the evolution of
Mercury's eccentricity, in the following we always consider all the
perturbers from Venus to Neptune. To simplify the notations, we define
two new polynomials $\Pb$ and $\Qb$ of the eccentricity $e_1$ alone,
given by (Eq. \ref{eq.Hamres0})
\be
\EQM{
\Pb(x) &= \frac{n_1}{g_5}
         \sum_{p=2}^8 \frac{m_p}{m_0} \frac{a_1}{a_p} 
         \Pa\left(\frac{a_1}{a_p}, x\right)\ ,
\crm 
\Qb(x) &= \frac{n_1}{g_5}
         \sum_{p=2}^8 \frac{m_p}{m_0} \frac{a_1}{a_p} 
         \Qa\left(\frac{a_1}{a_p}, x\right) \times A_p \ .
}
\ee
Then, the resonant Hamiltonian (\ref{eq.Hamres0}) reads as
\be
\check{H}_{N,{\rm plan}} = -g_5 \big(\Pb(e_1^2) - e_1 \Qb(e_1^2)
\cos(\varpi_1-\varpi_5^\star) \big) \ .
\label{eq.Hamres}
\ee
The numerical values of the coefficients of the polynomials $\Pb$ and
$\Qb$ are given in Appendix~\ref{sec.appnum}.

The Hamiltonian $\check{H}_{N,{\rm plan}}$ (\ref{eq.Hamres}) has one and a half
degrees of freedom, but it can be reduced to a {\em one} degree of
freedom Hamiltonian $\tilde{H}_{N,{\rm plan}}$ after \mybf{some
modifications}.
\mybf{
For that, we first make the Hamiltonian autonomous by a adding a
momentum $\check T$ conjugated to time $t$. Then, we perform a
canonical transformation where the old variables are
\be
(\check I = 1-\sqrt{1-e_1^2}\ ,  \check \theta = -\varpi_1)\ ;
\quad
(\check T\ ,t)\ ,
\ee
and the new ones are $(I, \Delta\varpi)$ and $(\tilde T, \tilde t)$,
defined by
\be
\EQM{
\Delta\varpi = -\check \theta - g_5 t - \varphi_5^\star \equiv \varpi_1 -
\varpi_5^\star\ , \cr
\tilde t = t\ .
}
\ee
For this transformation to be canonical, the momentums should verify
\be
\EQM{
\check I = - I \equiv 1-\sqrt{1-e_1^2}\ ,\cr
\check T = -g_5 I + \tilde T\ .
}
\ee
The new Hamiltonian $\tilde{H}_{N,{\rm plan}}$ expressed in the new
variables does not depend on the cyclic coordinate $\tilde t$. Thus, its
conjugated momentum $\tilde T$ is an integral of the motion and can be dropped.
Up to a constant, the new Hamiltonian is then
}
\be
\tilde{H}_{N,{\rm plan}} = -g_5\sqrt{1-e_1^2} -g_5  
\Pb\left(e_1^2\right) 
+ g_5 e_1 \Qb\left(e_1^2\right) \cos
\Delta\varpi\ .
\label{eq.Ham1}
\ee
This one degree of freedom Hamiltonian is integrable. The orbits are given by $\tilde{H}_{N,{\rm
plan}}=Cte$. \mybf{The temporal evolution of the eccentricity $e_1$ and 
of the resonant angle $\Delta\varpi$ are deduced from the canonical
equations of motion. This leads to}
\be
\EQM{
\frac{de_1}{dt} &= \frac{\sqrt{1-e_1^2}}{e_1}\Dron{\tilde{H}_{N,{\rm
plan}}}{\Delta\varpi}\ ,\crm
\frac{d\Delta\varpi}{dt} &=-\frac{\sqrt{1-e_1^2}}{e_1}\Dron{\tilde{H}_{N,{\rm
plan}}}{e_1}\ .
}
\label{eq.dt}
\ee

\subsection{General relativistic precession}
The secular effect of relativity is described by 
\citep[e.g.,][]{Touma_etal_MNRAS_2009},
\be
H_R = -g_r \frac{1}{\sqrt{1-e_1^2}}\ ,
\ee
where $g_r = 3 (G m_0)^2/(\Lambda_1 a_1^2 c^2)$, and $c$ is the speed of
light. In the case of Mercury, we have $g_r \approx 0.41$"/yr. The total
Hamiltonian $\tilde{H}_{R,{\rm plan}}=\tilde{H}_{N,{\rm plan}}+H_R$,
including the Newtonian interaction and general relativity, becomes
\be
\EQM{
\tilde{H}_{R,{\rm plan}} = & -g_5\Big(\sqrt{1-e_1^2}
+ \Pb\left(e_1^2\right)
- e_1 \Qb\left(e_1^2\right) \cos \Delta\varpi \Big)
\crm &
 -g_r \frac{1}{\sqrt{1-e_1^2}}\ .
}
\label{eq.Hamtot}
\ee

\subsection{Additional control term}
\mybf{
Using the  Newtonian Hamiltonian
(\ref{eq.Ham1}), or   the relativistic one (\ref{eq.Hamtot}), 
the system is far from  the $g_1-g_5$ resonance, 
in a configuration where the amplitude
of oscillation of the eccentricity is small (see next section). Indeed,
taking an order of expansion $n=50$, we get $g_1 = 5.54$"/yr in
the Newtonian case and $g_1 = 5.96$"/yr in the relativistic case,
whereas the resonance occurs in the vicinity of $g_1=g_5=4.25$"/yr.
This result is slightly different, 
but  representative of the present behavior of Mercury's
eccentricity. Indeed, the present value of $g_1$ for the Solar System 
with GR is $g_1 = 5.59$"/yr \citep{LaskRobu2004a}, the differences with the present model 
being due to the simplifications that are made here.
To recover a model that is dynamically close 
to the real Solar System, we add a correction 
to the Hamiltonian which changes the value of $g_1$ by an increment $\delta_g$.
In addition, owing to the slow chaotic diffusion in the
inner Solar System, Mercury's precession frequency $g_1$
can change quasi-randomly and come close to the value of $g_5$,
 then leading to high unstability 
 \citep{Laskar_Icarus_1990,Laskar_Icarus_2008,
Batygin_Laughlin_ApJ_2008,Laskar_Gastineau_Nature_2009}.
To explore the evolution of Mercury's
eccentricity for different values of $g_1$ around the resonant frequency
$g_5$, the above-mentioned correction is added to both
$\tilde{H}_{N,{\rm plan}}$ and 
$\tilde{H}_{R,{\rm plan}}$; i.e.,
\be
H_{N,{\rm plan}} = \tilde{H}_{N,{\rm plan}} + \frac{1}{2}\delta_g e_1^2
\label{eq.HamNplan}
\ee
and
\be
H_{R,{\rm plan}} = \tilde{H}_{R,{\rm plan}} + \frac{1}{2}\delta_g e_1^2\ .
\label{eq.HamRplan}
\ee
In Eqs.~(\ref{eq.HamNplan}) and (\ref{eq.HamRplan}), the factor $\delta_g$
controls the precession frequency $g_1$. For $\delta_g=0$, we recover 
the values obtained with $\tilde{H}_{N,{\rm plan}}$ and 
$\tilde{H}_{R,{\rm plan}}$, respectively. Otherwise the value of $g_1$
is incremented by $\delta_g$. Appendix~\ref{sec.appdg} provides the
values of $\delta_g$ that put the system in exact resonance for all
orders of expansion of the Hamiltonians.
}

\section{Eccentricity behavior}
\mybf{
Here, we analyze the possible trajectories and evolutions of Mercury's
eccentricity described by the Hamiltonians $H_{N,{\rm plan}}$
(\ref{eq.HamNplan}) and $H_{R,{\rm plan}}$ (\ref{eq.HamRplan}) obtained
in the previous section. Since these Hamiltonians are integrable (they
have only one degree of freedom), all the orbits are necessarily
regular. Thus, the goal of this section is not to reproduce the chaotic
behavior of Mercury's eccentricity $e_1$, but to show that $e_1$ can
actually reach values as high as observed by \citet{Laskar_Icarus_2008}
(see Fig.~\ref{fig.Comp}a) when the system is in the vicinity of the
resonance $g_1=g_5$. The chaotic evolution will be analyzed in the
spatial case (see Sect.~\ref{spatial}), where a new degree of freedom is
added.
}

\subsection{Dependency with the order of expansion}
\mybf{The Hamiltonians of the previous section have been computed as
series in power of the semimajor axis ratios $(a_1/a_p)_{p=2,8}$ for any
order $n$. Here, we study the effect of the truncation of these series.
Hereafter, we note with a superscript $(n)$ any Hamiltonian expanded up
to the order $n$, e.g., $H_{N,{\rm plan}}^{(n)}$ or $H_{R,{\rm plan}}^{(n)}$.
In a first step, we focus only on the relativistic Hamiltonian which is
more complete.
}

Within the quadrupole approximation, the Hamiltonian $H_{R,{\rm
plan}}^{(2)}$ reduces to a simple expression
\be
\EQM{
H_{R,{\rm plan}}^{(2)} = & -g_5 \left(\sqrt{1-e_1^2}
+\varepsilon^{\rm quad} \left(2+3e_1^2\right)\right) 
\crm &
- g_r \frac{1}{\sqrt{1-e_1^2}} + \frac{1}{2}\delta_g e_1^2\ ,
}
\label{eq.Ham2}
\ee
where $\varepsilon^{\rm quad} = \sum_p \varepsilon^{\rm quad}_p$.
This Hamiltonian is independent of $\Delta\varpi$, thus $e_1$ remains
constant. All the trajectories in the plane ($x=e_1\cos\Delta\varpi$,
$y=e_1\sin\Delta\varpi$) are circles centered on the origin (0,0) 
\mybf{with an eventually infinite period when $\Delta\dot\varpi=0$.
Figure~\ref{fig.section}a illustrates this result for $g_1=5$"/yr.
}
The position of the fixed points is represented by red dashed curves.
Within this approximation, as noted before, there is no possible
resonance between $\varpi_1$ and $\varpi_5^\star$ and thus, no possible
increase in the eccentricity.

The next level of approximation is the octupole. The corresponding
Hamiltonian reads as
\be
\EQM{
H_{R,{\rm plan}}^{(3)} =& -g_5 \bigg( \sqrt{1-e_1^2} 
               + \varepsilon^{\rm quad} (2+3e_1^2)
\cr &
               - \varepsilon^{\rm octu} (4+3e_1^2) \cos\Delta\varpi
               \bigg)
-g_r \frac{1}{\sqrt{1-e_1^2}} + \delta_g e_1^2\ ,
}
\ee
with $\varepsilon^{octu} = \sum_p \varepsilon^{\rm octu}_p$.
\mybf{In that case the phase space can be much more complicated with
five
fixed points and two sets of separatrices (see Fig.~\ref{fig.section}b
obtained with $g_1=5$"/yr). The stable fixed points are represented by
black dots, and the unstable ones are noted with crosses. The dashed
curves correspond to the separatrices. The positions and existence of 
the fixed points depend on the value of the precession frequency $g_1$.
This is depicted in the subpanel labeled $\alpha^3$ of Fig.~\ref{fig.fixpt}. The
curves represent the position on the $x$ axis of the fixed points of the
phase space as a function of $g_1$. The labels in roman numerals
qualifying each fixed point are identical to those in
Fig.~\ref{fig.section}.
}

\figComp

\mybf{
It is interesting to have a look at all of Fig.~\ref{fig.fixpt}.
Indeed, 
}
as the order $n$ of the expansion of the Hamiltonian $H_{R,{\rm
plan}}^{(n)}$ increases, the degrees of the polynomials $\Pa(\alpha,
e^2)$ and $\Qa(\alpha, e^2)$ increase both for $\alpha$ and $e^2$. Then,
the topology of the phase space of the Hamiltonian $H_{R,{\rm
plan}}^{(n)}$ evolves as shown in Fig.~\ref{fig.fixpt}. Up to $n=5$,
five fix points -- three stable points and two unstable points --
coexist within $ 4.5 \lesssim g_1 \lesssim 5.0$"/yr, while the system
contains at most three fix points -- two stable points and one unstable
point -- for $n \geq 6$. Moreover, the hyperbolic point (labeled IV)
located at $e_1 \cos \Delta\varpi > 0$, i.e. $\Delta\varpi=0$,
disappears when $n\geq 8$. The asymptotic topology is reached at
$n\approx 10$ from which positions of the fixed points do not evolve
significantly up to $n=20$.

The figures~\ref{fig.section}c, \ref{fig.section}d,
\ref{fig.section}e, and \ref{fig.section}f display the allowed
trajectories of Mercury's eccentricity for $n=20$ which is assumed to be
representative of the asymptotic behavior. The values of $g_1$ are 2.5,
3.6, 4.0, and 5.0"/yr, respectively. At this order $n=20$, orbits are
very similar to those of a simple pendulum \mybf{with a large resonant
island allowing for large increases in eccentricity beyond $e_1=0.8$.
The separatrix disappears at $g_1 \geq 3.84$"/yr but high eccentricity
excursions are still possible, to a smaller extent, since the elliptic
fixed point (V) remains significantly offset with respect to the
center of the phase space as long as $g_1\lesssim 4.5$"/yr. For
example, with $g_1=4.0$"/yr (Fig.~\ref{fig.section}e), a trajectory
starting within $e_1 \leq 0.2$ can reach $e_1\approx0.7$.}

\mybf{
To conclude, the asymptotic phase space, which must reproduce Mercury's
eccentricity behavior as faithfully as possible, is obtained for
$n\approx20$. Moreover, at this order of expansion, Mercury's
eccentricity is able to increase from $e_1\approx0.2$ up to $e_1\approx0.8$
as observed in the numerical simulations reported by
\citet{Laskar_Icarus_2008} (see Fig.~\ref{fig.Comp}).
}

\fignsection

\subsection{Temporal evolution}
The level curves of the Hamiltonian $H_{R,{\rm plan}}^{(20)}$ provide the
trajectories of Mercury's eccentricity in the phase space. In the
previous section, we saw that the eccentricity is able to vary between
0.2 and 0.8 as observed in \citep{Laskar_Icarus_2008} 
(Fig.~\ref{fig.Comp}a). We now check whether the timescale of the
evolution matches the 4 Myr found by \citet{Laskar_Icarus_2008}. For
this purpose, we integrate the following equations
\be
\EQM{
\frac{dx}{dt} = &\sqrt{1-e_1^2}\, \Dron{H_{R,{\rm plan}}^{(20)}}{y}\ ;\crm
\frac{dy}{dt} =-&\sqrt{1-e_1^2}\, \Dron{H_{R,{\rm plan}}^{(20)}}{x}\ ,
}
\label{eq.integ}
\ee
where $x=e_1\cos\Delta\varpi$ and $y=e_1\sin\Delta\varpi$. They
are equivalent to the equations of motion (\ref{eq.dt}),
but without singularities at the origin $e_1=0$.

An example of temporal evolution given by the numerical integration of
Eq.~(\ref{eq.integ}) is displayed in Fig.~\ref{fig.Comp}b. The initial
conditions are $e_1=0.2$, $\Delta\varpi=110\deg$, and $g_1=3.68$"/yr.
The comparison of Fig.~\ref{fig.Comp}a with Fig.~\ref{fig.Comp}b shows
that the simple \mybf{one degree of freedom} model described in this
study is able to reproduce the main evolution of Mercury's eccentricity
with the correct amplitude and timescale. Only once the eccentricity
reaches a value close to 0.8,  Mercury's orbit becomes highly unstable
due to close encounters with Venus, and the evolution observed in
\citep{Laskar_Icarus_2008} starts to differ significantly from the
simple model. It should be noted that the small oscillations present in
the integration of the full Solar System (Fig.~\ref{fig.Comp}a) do not
appear in the simple model since we keep only one single eigen frequency
to describe the evolution of the outer planets eccentricities.

\fignewt

\subsection{Without relativity}
We now consider the evolution of Mercury's eccentricity without
general relativity, i.e., described by the Hamiltonian $H_{N,{\rm
plan}}^{(n)}$ (\ref{eq.HamNplan}).
Figure \ref{fig.nsection} shows the level curves of $H_{N,{\rm
plan}}^{(20)}$ and $H_{N,{\rm plan}}^{(50)}$. The two figures are
similar to the subpanel $\alpha^{20}$ of Fig.~\ref{fig.section} within
the region of low eccentricities ($e_1 \lesssim 0.6$). But for higher
eccentricities, two new fixed points appear in the Newtonian case with
$n=20$. It is then necessary to extend the expansion up to the order
$n\approx50$ to get the asymptotic topology of the phase space. Once the
limit is closely reached, the result matches the relativistic ones well. The
only difference is in the value of $\delta_g$: for a given frequency
$g_1$, it should be increased by $g_r=0.41$"/yr with respect to the
relativistic case (see Table \ref{tabB2}).

Figure \ref{fig.newton} shows the positions of the fixed points and the
temporal evolution of the eccentricity driven by $H_{N,{\rm plan}}^{(50)}$.
There is no significant difference with respect to the relativistic
case.

\section{Spatial model}
\label{spatial}
The evolutions studied in the previous sections are perfectly regular.
Indeed, the Hamiltonian contains only one degree of freedom and is thus
integrable. \mybf{The goal was to show that it is possible to explain
the large increase in Mercury's eccentricity observed in numerical
simulations \citep[e.g.,][]{Laskar_Icarus_2008}. Our aim is now to
introduce an additional degree of freedom to make the system non
integrable and to reproduce a chaotic evolution expected in the vicinity
of the separatrix of the $g_1-g_5$ resonance.}

To do so, we add inclination in our system, and focus on the term
related to $(g_1-g_5) - (s_1-s_2)$. This resonant term is at present in
libration, and  was identified as one of the main source of chaos in the
Solar System \citep{Laskar_Icarus_1990}. It was indeed also computed
analytically in \citep{laskar_theorie_1984}, where it was identified as
the major obstacle for the convergence of the secular perturbation
series.  The Hamiltonian of the inclined problem is (see
Appendix~\ref{sec.appinc})
\be
\EQM{
\check{H}_{R,{\rm inc}} =&  
- g_r\frac{1}{\sqrt{1-e_1^2}}
-n_1 \sum_{p=2}^8 \bigg(\frac{m_p}{m_0}\left(\frac{a_1}{a_p}\right)^{n+1}
\crm & \times
{\cal F}_{n}^{(0,0)}
\left(e_1,A_p,i_1,I_p,\varpi_1-\varpi_5^\star,\Omega_1-\Omega_2^\star\right) \bigg)\ ,
}
\label{eq.Haminc0}
\ee
\mybf{where $i_p$ and $\Omega_p$ are the inclination and the longitude
of the ascending node of the planet $p$, and $I_p$ and $\Omega_2^\star$ are
the amplitude and the argument} of the term precessing at the frequency
$s_2$ in the quasiperiodic decomposition of
$\sin(i_p/2)\exp(\ii\Omega_p)$\mybf{, respectively. In
(\ref{eq.Haminc0}), both $\varpi_5^\star=g_5 t+\varphi_5^\star$ and 
$\Omega_2^\star=s_2 t+\phi_2^\star$ are functions of time $t$. To
reduce the number of degrees of freedom from 2.5 to 2, we apply
transformations similar to those of the planar case
(Sect.~\ref{planar}). We first add to the Hamiltonian the momentum 
$\check T$ conjugated to the time $t$. The old variables of the
subsequent canonical transformation are
\be
\EQM{
\check I = 1-\sqrt{1-e_1^2}\ ,\quad & 
\check \theta = -\varpi_1\ ,
\cr 
\check J = \sqrt{1-e_1^2}(1-\cos i_1)\ ,\quad &
\check \psi = -\Omega_1\ ,
\cr
\check T\ , & t\ ,
}
\label{eq.oldinc}
\ee
and the new variables are denoted $(I, \Delta\varpi)$, $(J,
\Delta\Omega)$, $(\tilde T, \tilde t)$. We set
\be
\EQM{
\Delta\varpi = -\check \theta - g_5 t - \varphi_5^\star 
\equiv \varpi_1-\varpi_5^\star\ ,\crm
\Delta\Omega = -\check \psi - s_2 t - \phi_2^\star 
\equiv \Omega_1 - \Omega_2^\star\ ,\crm
\tilde t = t\ .
}
\ee
The associated conjugated momenta are given by
\be
\EQM{
\check I = -I\ ,\crm
\check J = -J\ ,\crm
\check T = -g_5 I -s_2 J + \tilde T\ .
}
\ee
Since the new Hamiltonian $\tilde H_{R, {\rm inc}}$ is independent of
$t$, its conjugate momentum $\tilde T$ is an integral of the motion. 
From the expressions of $\check I$ and $\check J$ (\ref{eq.oldinc}), we
thus get, up to a constant,
\be
\EQM{
\tilde{H}_{R,{\rm inc}} &= 
\left(-g_5+2s_2\sin^2\frac{i_1}{2}\right)\sqrt{1-e_1^2}
-g_r \frac{1}{\sqrt{1-e_1^2}} \crm &
-n_1\sum_{p=2}^8 \frac{m_p}{m_0} \left(\frac{a_1}{a_p}\right)^{n+1}
{\cal
F}_n^{(0,0)}\left(e_1,A_p,i_1,I_p,\Delta\varpi,\Delta\Omega\right)\ .
}
\ee
To obtain the final Hamiltonian $H_{R,{\rm inc}}$ in the spatial case, we
add the control term and get
\be
H_{R,{\rm inc}} = \tilde{H}_{R,{\rm inc}} + \frac{1}{2}\delta_g e_1^2\ .
\label{eq.Haminc}
\ee
}
The Hamiltonian (\ref{eq.Haminc}) now has two degrees of freedom. The
variables are $(e_1, \Delta\varpi)$ and $(i_1, \Delta\Omega)$. To
perform numerical integrations, we  use
nonsingular rectangular coordinates
\be
\EQM{
&k=e_1\cos\Delta\varpi\ , \quad 
&h=e_1\sin\Delta\varpi \crm
&q=\sin\frac{i_1}{2}\cos\Delta\Omega\ , \quad 
&p=\sin\frac{i_1}{2}\sin\Delta\Omega\ .
}
\ee
With $\chi=\sqrt{1-e_1^2}$, the equations of motion read as \citep[e.g.][]{Bret1974a}
\be
\EQM{
\frac{dk}{dt} &= +\chi\Dron{H_{\rm inc}}{h} + \frac{h}{2\chi} 
\left(q\Dron{H_{\rm inc}}{q}+p\Dron{H_{\rm inc}}{p}\right)\ ,
\crm
\frac{dh}{dt} &= -\chi\Dron{H_{\rm inc}}{k} - \frac{k}{2\chi} 
\left(q\Dron{H_{\rm inc}}{q}+p\Dron{H_{\rm inc}}{p}\right)\ ,
\crm
\frac{dq}{dt} &= +\frac{1}{4\chi}\Dron{H_{\rm inc}}{p} -\frac{q}{2\chi}
\left(h\Dron{H_{\rm inc}}{k}-k\Dron{H_{\rm inc}}{h}\right)\ ,
\crm
\frac{dp}{dt} &= -\frac{1}{4\chi}\Dron{H_{\rm inc}}{q} -\frac{p}{2\chi}
\left(h\Dron{H_{\rm inc}}{k}-k\Dron{H_{\rm inc}}{h}\right)\ .
}
\ee

\figincli

Since Venus is the only planet with a significative amplitude
associated to the frequency $s_2$, it is also the only planet for
which the inclination is taken into account. The initial inclination of
Mercury is taken from \citep{Laskar_Icarus_1990} where all the terms in
its frequency decomposition, except the constant one, have been added.
Doing so, Mercury's initial inclination is measured with respect to the
invariant plane.

Figure \ref{fig.incli} shows the results of a numerical integration of
the inclined system. A large chaotic zone appears in the vicinity of the
 separatrix, as expected. Mercury's 
eccentricity now switches  randomly between low-amplitude 
circulations and high-amplitude excursions, reaching values that are close to 1.

\mybf{
It should ne noted that the Solar System is not presently in this chaotic zone. 
As described elsewhere \citep{Laskar_Icarus_1990,Laskar_Icarus_2008,
Batygin_Laughlin_ApJ_2008,Laskar_Gastineau_Nature_2009,Lithwick_Wu_ApJ_2011}, the system 
is at present in a secular resonance (with resonant argument $(g_1-g_5) - (s_1-s_2)$), 
in a state of slow chaos, with slow chaotic diffusion. 
This diffusion will quasi-randomly change the  value of $g_1$, which can then 
approach resonance with $g_5$. The phase diagram of Fig.~\ref{fig.incli} describes 
the behavior of Mercury's eccentricity once this slow diffusion has brought 
the system into the vicinity of the $g_1 -g_5$ resonance. 
}

\section{Conclusion}
In this study, we developed a simple model to account for the increase
in Mercury's eccentricity due to the resonance $g_1=g_5$. This  
\mybf{coplanar} model is based on the expansion of the perturbing function
with respect to the semimajor axis ratios, and exact in eccentricity.
In the resulting Hamiltonian, we kept only one term in the quasiperiodic
evolution of the outer planets' eccentricity.  We found that \mybf{this}
approximation is sufficient to reproduce an increase in Mercury's
eccentricity up to 0.8. But we also noticed that it is necessary to
extend the expansion up to the order $n\approx20$ and $n\approx 50$ in
the relativistic and in the Newtonian cases, respectively. The explicit
form of this secular resonant Hamiltonian is provided in Appendix
\ref{sec.appnum}.

The asymptotic topologies of the phase space are very similar no matter whether
the relativity is taken into account or not, so these two cases just
depend on the value of the resonant frequency (Table \ref{tabB2}). In
both cases, the system is a one degree of freedom system that is
integrable with a separatrix (Figs. \ref{fig.section}.d,
\ref{fig.nsection}).  The eccentricity of the trajectories in the
vicinity of this separatrix can rise to very high values, up to 0.8.
Moreover, the timescale of Mercury's evolution is in very good agreement
with the one observed in the numerical integration of the full model
\citep{Laskar_Icarus_2008} (Fig. \ref{fig.Comp}). With this integrable
model, the behavior of the numerical solutions computed in
\citep{Laskar_Icarus_2008,
Batygin_Laughlin_ApJ_2008,Laskar_Gastineau_Nature_2009} are understood,
but not the transition from a low-eccentricity regime to a high-eccentricity 
regime. To obtain such a transition, it is
necessary to include an additional degree of freedom in the system,
which transforms  the separatrix into a chaotic zone. 
 
 Since we know that the resonant term associated with $(g_1-g_5)-(s_1-s_2)$
has large amplitude \citep{laskar_theorie_1984,
Laskar_Icarus_1990,Lithwick_Wu_ApJ_2011}, we added this single term
in the spatial problem \mybf{which corresponds to our second model}
(Sec. \ref{spatial}). As expected, in this nonintegrable problem, the
separatrix is replaced by a significant chaotic zone where transitions
from low-eccentricity to high-eccentricity regimes occur quasi-randomly,
as observed in the full system, with maximal eccentricity close to 1
(Fig.\ref{fig.incli}).

\section*{Acknowledgments}
This work has been supported by PNP-CNRS, by the CS of the Paris Observatory,
by PICS05998 France-Portugal program, by the European Research Council/European
Community under the FP7 through a Starting Grant, as well as in the form
of grant reference PTDC/CTE-AST/098528/2008, funded by the Funda\c{c}\~ao
para a Ci\^encia e a Tecnologia (FCT), Portugal.

\appendix

\section{Explicit expression of 
the Hamiltonian development in the planar case}
\label{sec.appplan}
The secular Hamiltonian $\hat{H}_{\rm plan}$ of a restricted two-planet system where the 
massless body is on the inner orbit reads as
\be
\hat{H}_{\rm plan} = -\frac{G m_p}{a_p} \sum_{n=2}^\infty \alpha^n 
{\cal F}_n^{(0,0)} (e_1, e_p, \varpi_1-\varpi_p)\ .
\label{eq.AHam1}
\ee
The indices 1 and $p$ refer to the inner massless body and to the outer
massive planet, respectively; $m_p$ is the mass of the outer planet;
$\alpha=a_1/a_p$ is the semimajor axis ratio; $(e_k)_{k=1,p}$ and
$(\varpi_k)_{k=1,p}$ are the eccentricities and the longitudes of
periastron of the two planets.

In (\ref{eq.AHam1}), the ${\cal F}_n^{(0,0)} (e_1, e_p, \varpi)$ are 
given by \citep[see][]{Laskar_Boue_AA_2010}
\be
\EQM{
{\cal F}_n^{(0,0)} (e_1, e_p, \varpi) = \epsilon_n
f_{n,\frac{n}{2}} X_0^{n,0}(e_1) X_0^{-(n+1),0}(e_p) \crm
+\sum_{q=0}^{[(n-1)/2]} 2 f_{n,q} 
                        X_0^{n,n-2q}(e_1)
                        X_0^{-(n+1),n-2q}(e_p)
                        \cos((n-2q)\varpi)\ ,
}
\label{eq.AF00}
\ee
where $\epsilon_n=1$ if $n$ is even and 0, otherwise
\be
f_{n,q} = \frac{(2q)!(2n-2q)!}{2^{2n}(q!)^2((n-q)!)^2}\ ,
\label{eq.fnq}
\ee
and $X_0^{n,m}(e)$ are Hansen coefficients.

Now, we assume that the eccentricity of the outer planet remains low,
and we only keep the linear terms in $e_p$. Since the lowest power in
eccentricity of $X_0^{-(n+1),m}(e)$ is $e^{\abs{m}}$, all
the terms in the sum (\ref{eq.AF00}) are dropped, except those for which
$n-2q\leq1$.  Furthermore, from the explicit expressions of the Hansen
coefficients \citep{Laskar_Boue_AA_2010},
\be
\EQM{
X_0^{-n,m} = & \frac{1}{(1-e^2)^{n-3/2}} \crm
    & \times \sum_{\ell=0}^{[(n-2-m)/2]} \frac{(n-2)!}{\ell!(m+\ell)!(n-2-(m+2\ell))!}
      \left(\frac{e}{2}\right)^{m+2\ell} \,
}
\ee
for $n\geq2$, one gets
\be
\EQM{
X_0^{-n,0} = 1 + O(e^2)\ , \crm
X_0^{-n,1} = \frac{n-2}{2} e + O(e^3)\ .
}
\label{eq.ha}
\ee
Then, the expression of ${\cal F}_n^{(0,0)}$ simplifies as
\be
\EQM{
{\cal F}_n^{(0,0)} (e_1, e_p, \varpi) = & \epsilon_n
f_{n,\frac{n}{2}} X_0^{n,0}(e_1) \crm
& +(n-1)(1-\epsilon_n) f_{n,\frac{n-1}{2}} 
                        X_0^{n,1}(e_1) e_p
                        \cos\varpi \crm
& + O(e_p^2) \ .
}
\ee
Substituting this expression into (\ref{eq.AHam1}), one obtains
\be
\EQM{
\hat{H}_{\rm plan} = & -\frac{G m_p}{a_p} \Bigg(
    \Pa\left(\alpha, e_1^2\right) 
  - e_1 e_p \Qa\left(\alpha, e_1^2\right) \cos (\varpi_1-\varpi_p)
  \Bigg) 
\crm &
+ O(e_p^2)\ ,
}
\label{eq.AHam3}
\ee
where
\be
\Pa\left(\alpha, e^2 \right) = \sum_{q=1}^\infty \alpha^{2q} f_{2q,q}
X_0^{2q,0} (e)\ ,
\ee
and
\be
\Qa\left(\alpha, e^2 \right) = -\frac{1}{e}\sum_{q=1}^{\infty} 2q \alpha^{2q+1} f_{2q+1,q}
X_0^{2q+1,1}(e) \ .
\ee
To derive explicit expressions of these two quantities, we use the
analytical formulae of the coefficients $X_0^{n,m}(e)$ for $n\geq 2$
\citep{Laskar_Boue_AA_2010},
\be
\EQM{
X_0^{n,m} (e) = & (-1)^m \frac{(n+1-m)!}{(n+1)!} \crm &
  \times \sum_{\ell=0}^{[(n+1-m)/2]} 
         \frac{(n+1-m)!}{\ell!(m+\ell)!(n+1-m-2\ell)!} 
  \left(\frac{e}{2}\right)^{m+2\ell}\ .
}
\ee
Moreover, we invert the sums and separate the cases $\ell=0$ and
$\ell \geq 1$ in $\Pa(\alpha, x)$. This gives
\be
\EQM{
\Pa(\alpha, x) & = \sum_{q=1}^{\infty} f_{2q,q} \alpha^{2q}
    \crm &
    + \sum_{\ell=1}^{\infty} \left(
      \sum_{q=\ell}^\infty f_{2q,q} \frac{(2q+1)!}{\ell!\ell!(2q+1-2\ell)!}
      \alpha^{2q} \right) \left(\frac{x}{4}\right)^\ell\ ,
}
\label{eq.Pa}
\ee
and
\be
\EQM{
\Qa(\alpha, x) =
      \sum_{\ell=0}^{\infty} \Bigg( &
      \sum_{q=\ell}^\infty f_{2q+1,q} 
      \crm &
      \times\frac{q(2q+3) (2q+1)!}{\ell!(\ell+1)!(2q+1-2\ell)!}
      \alpha^{2q+1} \Bigg) \left(\frac{x}{4}\right)^\ell \ .
}
\label{eq.Qa}
\ee
It should be noted that in (\ref{eq.Pa}) and (\ref{eq.Qa}), the upper
limits of $\ell$ and $q$ are infinite because we consider here the
infinite expansion of the perturbing function in series of $\alpha$.
Nevertheless, when the Hamiltonian is truncated at the order $\alpha^n$,
the sums become finite. In $\Pa\,$, the maximum value taken by $\ell$ and
$q$ is $[n/2]$, and in $\Qa\,$, it is $[(n-1)/2]$.

\section{Numerical coefficients of the polynomials $\Pb$
and $\Qb$}
\label{sec.appnum}

Here, we consider the system composed of Mercury perturbed by the seven
outer planets from Venus to Neptune ($p=2,8$). The secular Hamiltonian
of this problem, truncated at the first order in the outer
eccentricities reads as
\be
\EQM{
\moy{M_p}{\hat{H}_{N,{\rm plan}}} 
   = -\sum_{p=2}^8 & \frac{G m_p}{a_p} \Bigg(
    \Pa\left(\frac{a_1}{a_p}, e_1^2\right) 
\crm &
  - e_1 e_p \Qa\left(\frac{a_1}{a_p}, e_1^2\right) \cos (\varpi_1-\varpi_p)
  \Bigg)\ .
}
\label{eq.BHam1}
\ee
Then, using the quasiperiodic expansion of the variables
$(e_p\e^{i\varpi_p})_{p=2,8}$ and keeping only the terms at the fundamental
frequency $g_5$ $(A_p\e^{i\varpi^\star_5})_{p=2,8}$, one gets
\be
\EQM{
\moy{M_p}{\hat{H}_{N,{\rm plan}}}
 = -\sum_{p=2}^8 & \frac{G m_p}{a_p} \Bigg(
    \Pa\left(\frac{a_1}{a_p}, e_1^2\right) 
\crm &
  - e_1 A_p \Qa\left(\frac{a_1}{a_p}, e_1^2\right) \cos (\varpi_1-\varpi^\star_5)
  \Bigg)\ .
}
\label{eq.BHam2}
\ee
We now define the polynomials $\Pb$ and $\Qb$ such that
\be
\EQM{
\check{H}_{N,{\rm plan}} & \equiv \moy{M_p}{\hat{H}_{N,{\rm plan}}}/\Lambda_1 
\crm &
= -g_5 \Pb\left(e_1^2\right) 
+ g_5 \Qb\left(e_1^2\right) e_1 \cos(\varpi_1-\varpi^\star_5)\ ,
}
\label{eq.BHam3}
\ee
where $\Lambda_1 = \sqrt{G m_0 a_1}$.
The new polynomials $\Pb$ and $\Qb$ are derived from $\Pa\,$ and $\Qa\,$
through
\be
\Pb(x) = \frac{n_1}{g_5}
         \sum_{p=2}^8 \frac{m_p}{m_0} \frac{a_1}{a_p} 
         \Pa\left(\frac{a_1}{a_p}, x\right)\ ,
\label{eq.Pbnum}
\ee
and
\be
\Qb(x) = \frac{n_1}{g_5}
         \sum_{p=2}^8 \frac{m_p}{m_0} \frac{a_1}{a_p} 
         \Qa\left(\frac{a_1}{a_p}, x\right) \times A_p\ .
\label{eq.Qbnum}
\ee
Their numerical values, summarized in Table \ref{tab.B1}, have been
computed up to the order $\alpha^{50}$ from \citep{Laskar_Icarus_1990}.
 
\tabBa

\section{Frequencies at zero eccentricity and corrections}
\label{sec.appdg}
The precession frequency $g_1$ of Mercury's perihelia computed
either from the Newtonian Hamiltonian (\ref{eq.Ham1}),
\be
\tilde{H}_{N,{\rm plan}} = -g_5\sqrt{1-e_1^2} -g_5  
\Pb\left(e_1^2\right) 
+ g_5 e_1 \Qb\left(e_1^2\right) \cos
\Delta\varpi\ ,
\ee
or from the relativistic Hamiltonian 
(\ref{eq.Hamtot})
\be
\EQM{
\tilde{H}_{R,{\rm plan}} = & -g_5\Big(\sqrt{1-e_1^2}
+ \Pb\left(e_1^2\right)
- e_1 \Qb\left(e_1^2\right) \cos \Delta\varpi \Big)
\crm &
 -g_r \frac{1}{\sqrt{1-e_1^2}}
}
\ee
is far enough from $g_5\approx4.249$"/yr for the system not to be in
secular resonance (see Table~\ref{tabB2}). Nevertheless, due to the slow
diffusion of the inner planets of the Solar System, this frequency is
subject to small variations, and Mercury can eventually reach the
resonance. To model this change in frequency, we include an additional
term in the Hamiltonians: $\delta_g e_1^2/2$. 

\tabBb

The values of $\delta_g$ putting the system in exact resonance at zero
eccentricity are shown in Table~\ref{tabB2}. One can observe that the
correction is higher (in absolute value) when the relativistic
precession is taken into account, which explains why relativity
stabilizes the Solar System.

\section{Explicit expression of the Hamiltonian development
in the spatial case}
\label{sec.appinc}
Here we develop the secular Hamiltonian of an inclined restricted
two-planet system following \citet{Laskar_Boue_AA_2010}. We use the same kind of approximation as for the
planar case by considering only the linear dependency on the
eccentricities and inclinations of the outer planets. 
We note $i_1$ and $i_p$ the inclinations of the massless planet and of the
perturber respectively, and $\Omega_1$ and $\Omega_p$ are their longitudes of
ascending node. We also note $c_1=\cos{(i_1/2)}$, $s_1=\sin{(i_1/2)}$,
$c_p=\cos{(i_p/2)}$, and $s_p=\sin{(i_p/2)}$, from which we define
\be
\EQM{
\mu_* &= \left(c_1c_p\e^{\ii\frac{\Omega_1-\Omega_p}{2}}
         +s_1s_p\e^{-\ii\frac{\Omega_1-\Omega_p}{2}}\right)^2\ , \crm
\nu_* &= \left(c_1s_p\e^{\ii\frac{\Omega_1-\Omega_p}{2}}
         -s_1c_p\e^{-\ii\frac{\Omega_1-\Omega_p}{2}}\right)^2\ .
}
\label{eq.munu}
\ee

According to \citet[][Eq. (B.30)]{Laskar_Boue_AA_2010}, the term in
factor of $\alpha^n$ in the perturbing function reads as
\be
\EQM{
{\cal F}_n^{(0,0)} = \sum_{s=0}^n\sum_{q=0}^n & \Big( \tilde
Q_{s,q}^{(n)}(\mu_*,\nu_*) X_0^{n,n-2s}(e_1) X_0^{-(n+1),n-2q}(e_p)
\crm & \times
\e^{\ii(n-2s)\omega_1} \e^{\ii(n-2q)\omega_p} \Big)\ ,
}
\label{eq.F00i}
\ee
Since we only consider the harmonics $(\varpi_1-\varpi_p)$ and
$((\varpi_1-\varpi_p) - (\Omega_1-\Omega_p))$, we keep the terms such
that $q=n-s$, and drop the others. Consequently, one sum disappears from
(\ref{eq.F00i}), and it remains 
\be
\EQM{
{\cal F}_n^{(0,0)} = \sum_{s=0}^n & \Big( 
\tilde Q_{s,n-s}^{(n)}(\mu_*,\nu_*) 
X_0^{n,n-2s}(e_1) X_0^{-(n+1),n-2s}(e_p) 
\crm & \times
\e^{\ii(n-2s)\Delta\omega} \Big)\ .
}
\ee
Then, we extract the linear terms in $e_p$. Since
$X_0^{-(n+1),m}(e_p)\propto e_p^{\abs{m}}$, we consider only the values
of $s$ such that $n-2s=0,\pm 1$. Using the asymptotic expressions of the
Hansen coefficients (\ref{eq.ha}), we get
\be
{\cal F}_{2n}^{(0,0)} = \tilde Q_{n,n}^{(2n)}(\mu_*,\nu_*) X_0^{2n,0}(e_1)\ ,
\label{eq.F2n}
\ee
and
\be
{\cal F}_{2n+1}^{(0,0)} = \tilde Q_{n,n+1}^{(2n+1)}(\mu_*,\nu_*) 
\,X_0^{2n+1,1}(e_1) \,n e_p \,\e^{\ii\Delta\omega} + cc\ ,
\label{eq.F2n1}
\ee
where $cc$ means complex conjugate.
Then, we use the definition of the $\tilde Q$ functions
\citep[][Eq. (B.31)]{Laskar_Boue_AA_2010}. For $s+q\leq n$, one has
\be
\tilde Q_{s,q}^{(n)} (\mu_*,\nu_*) = \mu_*^{q-s} \nu_*^{n-q-s}
A_{q-s,n-q-s}^{(n)}(\abs{\nu_*})\ ,
\ee
with \citep[][eq. (B.32)]{Laskar_Boue_AA_2010},
\be
\EQM{
A_{q-s,n-q-s}^{(n)} (x) = & \frac{1}{2^{2n}}
\frac{(2s)!(2n-2q)!}{s!(n-s)!q!(n-q)!}
\crm & \times
\sum_{k=0}^{2s} (-1)^k
\frac{(2n-2s+k)!}{(2s-k)!(2n-2q-2s+k)!}
\frac{x^k}{k!} \ . 
}
\ee
We thus have
\be
\EQM{
\tilde Q_{n,n}^{(2n)}(\mu_*,\nu_*) &= A_{0,0}^{2n}(\abs{\nu_*})
\crm &=
\frac{1}{2^{4n}}\frac{(2n)!(2n)!}{(n!)^4}
\sum_{k=0}^{(2n)} (-1)^k \frac{(2n+k)!}{(2n-k)!(k!)^2} \abs{\nu_*}^k\ ,
}
\ee
or in a more condensed form,
\be
\tilde Q_{n,n}^{(2n)}(\mu_*,\nu_*) = f_{2n,n} F(2n+1,-2n,1;\abs{\nu_*} )\ ,
\label{eq.Qnn}
\ee
where $F$ is the hypergeometric function and $f_{p,q}$ is given in
(\ref{eq.fnq}). In the same way,
\be
\tilde Q_{n,n+1}^{(2n+1)} = \mu_* f_{2n+1,n}
F(2n+3,-2n,1;\abs{\nu_*} )\ .
\label{eq.Qnn1}
\ee
Now, we substitute the expressions of $\mu_*$ and $\nu_*$
(\ref{eq.munu}) into those of $\tilde Q_{n,n}^{2n}$ (\ref{eq.Qnn}) and
of $\tilde Q_{n,n+1}^{2n+1}$ (\ref{eq.Qnn1}). Keeping only the linear
terms in inclinations, 
\be
\mu_* \approx c_1^2 \e^{\ii\Delta\Omega} + 2 c_1 s_1 s_p\ ,
\label{eq.mu*}
\ee
and
\be
\abs{\nu_*}^k \approx s_1^{2k} - 2 k c_1 s_1^{2k-1} s_p \cos\Delta\Omega\ ,
\label{eq.nu*}
\ee
with $\Delta\Omega=\Omega_1-\Omega_p$, and retaining only the terms involved in 
the secular resonances, we get 
\be
\EQM{
{\cal F}_{2n}^{(0,0)} &=& f_{2n,n} F(2n+1,-2n,1;s_1^2) X_0^{2n,0}(e_1)\ ;
\crm 
{\cal F}_{2n+1}^{(0,0)} &=& 2 f_{2n+1,n} \Big(
c_1^2 F(2n+3,-2n,1;s_1^2) \cos\Delta\varpi
\crm && +
c_1 s_1 s_p G_{2n+1,n}(s_1^2) \cos(\Delta\varpi-\Delta\Omega)
\Big)\,
X_0^{2n+1,1}(e_1)\,ne_p
\ ,
}
\label{eq.Finc}
\ee
where
\be
\EQM{
G_{2n+1,n}(x) =& 2 F(2n+3,-2n,1;x)
\crm &
               -(1-x)\,F'(2n+3,-2n,1;x)\ ,
}
\ee
and
\be
F'(a,b,c;x) = \frac{ab}{c}F(a+1,b+1,c+1;x)\ .
\ee
The Eqs.~(\ref{eq.Finc}) generalize the expression (\ref{eq.Fn00})
obtained in the coplanar problem. The Hamiltonian of the inclined
system is then
\be
\hat{H}_{\rm inc} = -\frac{G m_p}{a_p} \sum_{n=2}^\infty \alpha^n 
{\cal F}_n^{(0,0)} (e_1, e_p, i_i, i_p, \varpi_1-\varpi_p)\ ,
\label{eq.CHam}
\ee
with ${\cal F}_n^{(0,0)}$ given in (\ref{eq.Finc}).

\bibliographystyle{aa}
\bibliography{mercure}

\end{document}